\documentclass[%
reprint,
 amsmath,amssymb,
 aps,
prb,
]{revtex4-2}

\usepackage{graphicx}
\usepackage{dcolumn}
\usepackage{bm}
\usepackage{color}

\usepackage{float}
\usepackage{ulem}

\usepackage[dvipsnames]{xcolor}

\begin{document}
\preprint{APS/123-QED}

\title{Electron-phonon coupling in correlated materials: \\ insights from the Hubbard-Holstein model}
\author{Jennifer Coulter}
\affiliation{Center for Computational Quantum Physics, The Flatiron Institute,162 5th Avenue,New York, NY 10010}
\author{Andrew J. Millis}
\affiliation{Center for Computational Quantum Physics, The Flatiron Institute,162 5th Avenue,New York, NY 10010}
\affiliation{Department of Physics, Columbia University, 538 West 120th Street, New York, New York 10027}

\date{\today}

\begin{abstract}
Dynamical mean-field theory computations of the electron self energy of the Hubbard-Holstein model as a function of electron-phonon and electron-electron interactions are analyzed to gain insight into the dependence of electron-phonon couplings on correlation strength in quantum materials. We find that the electron-phonon interaction is strongly suppressed by electronic correlations, while electron-electron correlation effects at Fermi liquid scales are only weakly modified by coupling to phonons, with phonon-induced modifications most evident at high frequencies on the order of the electronic bandwidth. Implications for beyond-density functional theories of the electron-phonon interaction are discussed.  
\end{abstract}

\maketitle

\section{\label{sec:introduction}Introduction}

The coupling of electrons to quantized lattice vibrations (phonons) plays a fundamental role in the physics of solids~\cite{ziman2001electrons,giustino2017electron}. In many conventional materials, this coupling is adequately determined from density functional theory (DFT), 
in which the electron-phonon coupling is described by scattering amplitudes calculated from the change of the Kohn-Sham potential due to static atomic displacements away from equilibrium positions. One efficient implementation is through density functional perturbation theory within the framework of DFT~\cite{giustino2017electron}. These DFT matrix elements have been used to successfully predict scattering rates, spectral functions, polaronic properties, transport and optical effects, and BCS superconducting transition temperatures for many weakly correlated materials~\cite{giustino2017electron}. 

However, in many ``quantum" materials, the electron-phonon coupling strength is apparently strongly renormalized by beyond-DFT electronic correlations, being in some cases greatly underestimated~\cite{mandal2014strong,gerber2017femtosecond,yin2013correlation,gwpt_li_2019,abramovitch2023combining,abramovitch2024respective,abramovitch2025electron, lazzeri2008impact,chen2025impact, coulter2025mechanisms} and in others overestimated~\cite{gunnarsson2008interplay,abramovitch2025electron}. Further, whether the coupling constant renormalization is the only effect occurring in the correlated electron plus phonon problem is not known. Understanding these issues and translating the understanding to a practical computational scheme for correlated materials is an important area in materials theory. 

In order to make progress towards the goal of ab-initio electron-phonon coupling including correlation effects, it is important to use exactly solvable models to extract insights that will inform more realistic materials-based work. One approach is to apply the DFT methodology for electron-phonon coupling in the context of DFT+U, which neglects dynamical correlations. Another is based on the Hedin formalism, for which a theoretical approach has been outlined~\cite{giustino2017electron,gwpt_li_2019} and implemented in the leading order $GW$ approximation~\cite{li2024electron} under the approximation that changes in ``W" (representing the correlation strength) are typically neglected. While these studies serve as a demonstration of the importance of correlation effects in electron-phonon coupling~\cite{gwpt_li_2019, lazzeri2008impact, yin2013correlation}, workable extensions for more strongly correlated materials have not been established. 

To gain insight into the physics of electron-phonon coupling in correlated electron systems, in this paper we study the Hubbard-Holstein model~\cite{holstein1959studies}, in which electrons interacting via an on-site interaction are coupled to a dispersionless (single frequency) optical phonon mode. While the Holstein-Hubbard model has been extensively studied~\cite{capone2010electron_HH_review}, much of the focus has been on phase diagrams, in particular superconducting, density wave and bipolaron instabilities~\cite{Karakazu22,Han20,Ohgoe17,Johnson13,Costa20,Zhang23,becca1996charge_HH,sangiovanni_electron-phonon_2005}. The electron-phonon coupling has been studied using sum rules to evaluate overall coupling strengths \cite{rosch2004apparent,gunnarsson2008interplay}. Some aspects of electron-phonon coupling have been inferred from studies of the total electron mass renormalization~\cite{sangiovanni2006electron}, and analysis of the consequences of the compressibility singularities occurring near the Mott transition~\cite{moghadas2025effective}.  Here, we show that it is possible to analyze the electron-phonon coupling along lines similar to the conventional weak coupling theory, by isolating and investigating a specifically phonon contribution to the electron scattering rate and mass enhancement as a function of correlation strength and electron-phonon coupling. Our approach is related to previous work of Huang, Hanke, Arrigoni and Scalapino~\cite{Huang03} who used Monte Carlo methods to evaluate a formal diagrammatic expression for the interaction-induced renormalization of the electron-phonon coupling, see also \cite{Cappelluti04}. %

We employ the single-site dynamical mean-field approximation~\cite{georges1996dynamical}, along with a canonical transformation method~\cite{werner2007efficient} that enables a numerically exact solution of the full Holstein-Hubbard model including dynamical phonon effects. We compute the electron self energy as a function of electron-phonon and electron-electron interactions and isolate the effects of the electron-phonon interaction via a comparison of the self energies calculated with and without phonons at fixed values of the electron-electron interactions. Unlike standard approaches, our work does not make use of perturbation theory of a Migdal-Eliashberg approximation.
From the comparison we show that, as a first approximation, phonon effects are additive to electron-electron contributions and may be treated perturbatively (as in the standard theory), but with an effective electron-phonon coupling constant that is strongly renormalized by the correlations. Changes to correlation effects in the low-frequency Fermi liquid regime are minimal; however there are changes which appear at higher frequencies on the order of the bandwidth.

This paper proceeds by first presenting in Section~\ref{sec:Methods} the model and DMFT methodology. Then, in Section~\ref{sec:results}, we summarize the standard theory of electron-phonon interactions and present the DMFT results for self energies. We next discuss the renormalization of the electron-phonon coupling strength and phonon frequency in Section~\ref{sec:ph_renorm} and then the electron correlation strength in Section~\ref{sec:correlation_renorm}. We close with a summary and conclusions in Section~\ref{sec:conclusion}, with extended data shown in the SI. 
\section{\label{sec:Methods}Methods}
The Hubbard-Holstein model of interacting electrons moving on sites of a lattice coupled to dispersionless Einstein phonons is defined by the Hamiltonian,
\begin{eqnarray}
H_{HH}&=&\sum_{ij\sigma}t^{ij}c^\dagger_{i\sigma}c_{j\sigma}+U\sum_i\hat{n}_{i\uparrow}\hat{n}_{i\downarrow} \nonumber \\
&&+g\sum_{i}\left(b^\dagger_i+b_i\right)\hat{n}_i+\omega_0\sum_ib^\dagger_ib_i.
\label{eq:HHH} 
\end{eqnarray}
Here $c^\dagger_{i\sigma}$ creates an electron of spin $\sigma$ on site $i$, $\hat{n}_{i\sigma}=c^\dagger_{i\sigma}c_{i\sigma}$ gives the density of electrons of spin $\sigma$ on site $i$, $\hat{n}_i=\hat{n}_{i\uparrow}+\hat{n}_{i\downarrow}$ is the total density of electrons on site $i$, $U$ is the Hubbard interaction which we will take to be repulsive, $b^\dagger_i$ is the operator creating a phonon on site $i$ and $\omega_0$ is the bare phonon frequency. %

One may view the hopping amplitudes $t^{ij}$ as arising from the projection of the Kohn-Sham Hamiltonian of band theory onto a single Wannierizable band, the $b^\dagger$ as the creation operators for a weakly dispersing optical breathing mode, and $g$ as the projection of the DFT electron-phonon coupling onto the band of interest. The Hubbard $U$ expresses the beyond-DFT correlations.

We solve the model using the single-site dynamical mean-field approximation (DMFT), which approximates the full electronic self energy as a local (momentum-independent) function $\Sigma(\omega)$ that is determined from the solution of a quantum impurity model specified by a self-consistency equation~\cite{georges1996dynamical}. In DMFT, the only relevant feature of the hopping $t^{ij}$ is the density of states, $N$, which we take to have the semicircular form with bandwidth $4t$ so that $N(\varepsilon)=\frac{1}{\pi t^2}\sqrt{4t^2-\varepsilon^2}$. We treat the electron-phonon coupling via the method introduced in~\cite{werner2007efficient}, in which a canonical transformation eliminates the explicit electron-phonon coupling in favor of a phonon operator-dependent hybridization, which may be treated by a Feynman disentangling, allowing study of the full non-perturbative and dynamical phonon effects in the Hubbard-Holstein model at almost the same computational cost as the simple Hubbard model. 

To solve the auxiliary impurity model with the operator-dependent hybridization we used the CTSEG solver~\cite{werner2006continuous} as implemented in the TRIQS software ecosystem~\cite{parcollet2015triqs,Kavokine2025} using the improved Monte Carlo estimators of Ref.~\cite{improved_estimators}. 
The solver works at non-zero temperature and produces Matsubara-axis electron Green functions and self energies; to obtain real frequency results we performed analytic continuation using the Pad\'e approximant method of Ref.~\cite{vidberg1977solving} as implemented in the TRIQS package~\cite{parcollet2015triqs}. We also used cubic spline interpolation of Matsubara axis data to estimate zero frequency scattering rates. All analytical continuations were required to reproduce the imaginary axis data to within $0.1\%$ relative error for Matsubara frequencies $|\omega_n|$ less than $2t$; cross checks with cubic spline interpolations and systematics of dependence on parameters suggests that the continuations are reliable for frequencies $\lesssim t$; for higher frequencies the continuations may be less reliable. To compute the change in self energies due to the electron-phonon interaction, we continued the two self energies separately and took the difference in real frequency.

For this study, we restricted attention to temperatures $\geq t/100$. We consider two phonon frequencies: $\omega_0=0.2t$, which is small relative to the bandwidth but large enough that computations at $T=t/100$ are representative of the $T=0$ limit and $\omega_0=0.02t$, for which temperatures greater than the phonon frequency but still much less than the bandwidth could be studied.

\section{Results \label{sec:results}}
\subsection{Overview of electron-phonon effects, $U=0$ }
We begin by summarizing the standard result for the electron-phonon self energy of the uncorrelated $(U=0)$ system to establish notations and baseline physics. We focus on the most physically relevant case of phonon frequency $\omega_0$ small compared with electronic bandwidth. We will be interested both in the low temperature limit $T\ll \omega_0$ and the intermediate temperature regime $T\gtrsim \omega_0$ but still much less than the bandwidth. 

At $U/t=0$ and small $g$, the effects of the electron-phonon coupling may be computed perturbatively. At temperature $T=0$, the perturbative result for the imaginary part of the electron-phonon contribution to the self energy is,
\begin{equation}
\mathrm{Im}\left[\Sigma(\omega>0)\right] = \pi g^2 N(\omega-\omega_0) \Theta(\omega - \omega_0),
\label{eq:imsigmau0}
\end{equation}
with $N$ the density of states defined above. 
In the physically relevant limit of phonon frequency small compared to the electronic bandwidth, Im$\Sigma(\omega>0)=\frac{\pi}{2}\lambda\omega_0\Theta(\omega-\omega_0)$ for frequencies small compared to the electronic bandwidth, and goes to zero for frequencies larger than the bandwidth. Here  the dimensionless electron-phonon coupling $\lambda$ is defined as,
\begin{equation}\lambda=\frac{2g^2N(\mu)}{\omega_0}.
\label{eq:lambdadef}
\end{equation}

At low frequencies $|\omega| \ll\omega_0$, Re$\Sigma(\omega)\approx-\lambda \omega$. As $\omega$ increases -Re$\Sigma$ reaches a maximum at $\omega=\omega_0$ and then decreases, changing sign at a frequency 
that, for $\omega_0$ small relative to the bandwidth, is of the order of the geometric mean of the phonon frequency and bandwidth. Re$\Sigma$ then passes through a minimum at frequencies of order of the bandwidth and then tends to zero as frequency is further increased. For the semicircular density of states used here, in the limit $\omega_0\ll t$ and still at $T=0$,
\begin{multline} 
        \mathrm{Re}\left[\Sigma(\omega)\right]\approx -\frac{\lambda\omega_0}{2}\Biggl[ln\left|\frac{\omega+\omega_0}{\omega-\omega_0}\right|- 
        \\
        \pi \left(\frac{\omega}{2t}+sign(\omega)\Theta(\omega^2-4t^2)\sqrt{\frac{\omega^2}{4t^2}-1}\right)\Biggl].
    \label{eq:resigmau0}
\end{multline}
The formula yields a sign change at $\omega\approx \sqrt{\frac{4\omega_0t}{\pi}}$. 

As the temperature is increased from $T=0$, the imaginary part of the self energy becomes non-zero at all frequencies, the step at $\omega_0$ broadens and decreases in amplitude, and the low frequency peak in Re$\Sigma$ becomes washed out. For temperatures $\gtrsim \omega_0/2$, the zero frequency limit of the self energy is approximately linear in temperature, $\Sigma(\omega=0,T)=\pi\lambda T$.

As the dimensionless electron-phonon coupling $\lambda$ is increased beyond the perturbative regime ($\lambda \lesssim 0.1$ for the model considered here) two important nonlinear effects contribute. First, the electron-induced renormalization of the phonon propagator reduces the effective renormalized phonon frequency to $\tilde{\omega}_0\approx \omega_0(1-\lambda)$. Second, the amplitude of the effects increases nonlinearly with $\lambda$. For large enough $\lambda$, the ground state reconstructs to a charge density wave or bipolaron state, which is outside the scope of the current work. 

\subsection{Self energies for $U/t=0$}
\label{sec:self-energies-U0}
\begin{figure*}
        \centering
        \includegraphics[width=0.97\linewidth]{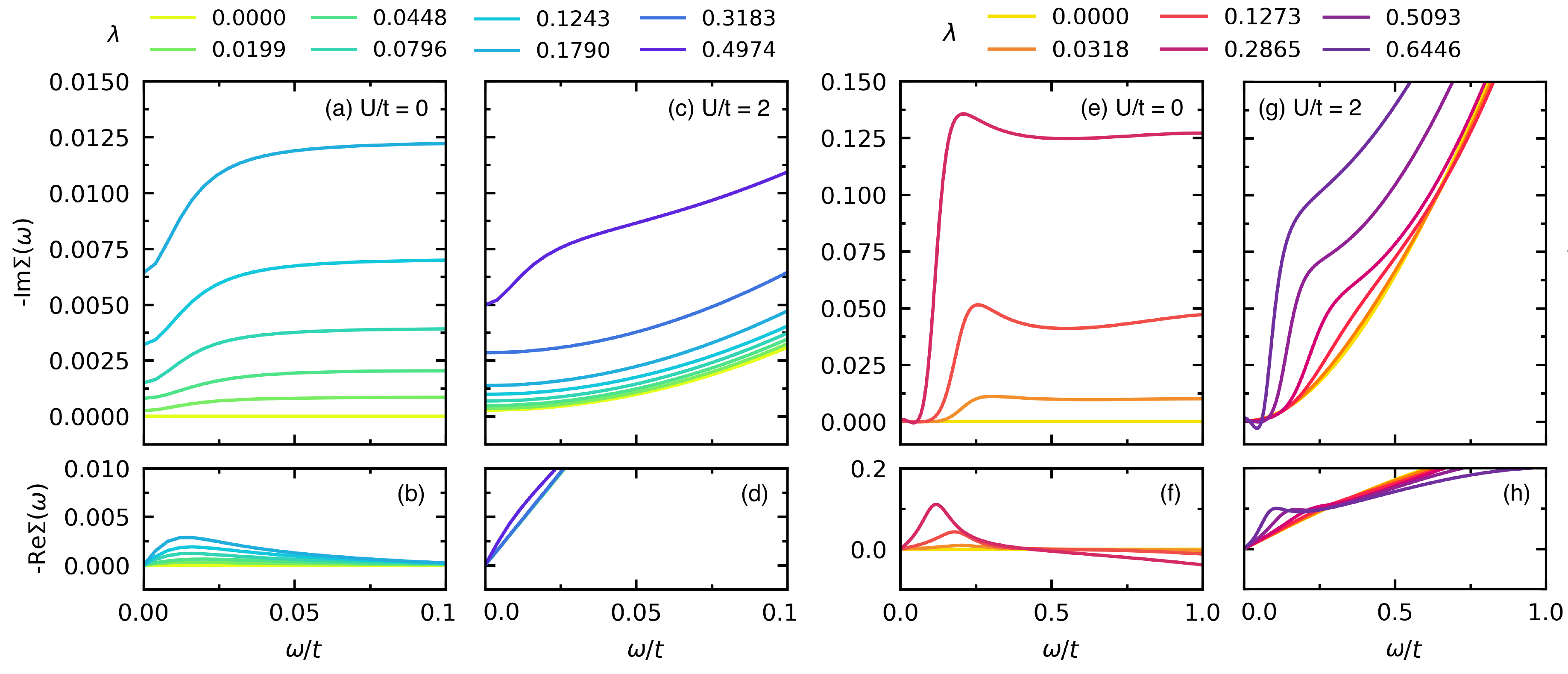}
        \caption{ \textbf{Electron self energies including phonon effects.} Imaginary ($a,c,e,g$) and real ($b,d,f,h$) parts of the analytically continued self energies for ($a,b,e,f$) $U/t$ = 0 and ($c,d,g,h$) $U/t$ = 2, calculated using a bare phonon frequency of $\omega_0/t$ = 0.02 for ($a-d$) and $\omega_0/t$ = 0.2 in ($e-h$), for a range of electron-phonon coupling $\lambda$ at a temperature of $\beta\cdot t$ = 100.}
        \label{fig:self_energies}
\end{figure*}

Panels ($a,b,e,f$) of Fig.~\ref{fig:self_energies} show the imaginary ($a,e$) and real ($b,f$) parts of the analytically continued electron self energy computed for $U/t=0$ ($a,b,e,f$) for both $\omega_0/t=0.02$ ($a,b$) (left panels in yellow to blue colors) and $0.2$ ($e,f$) (right panels in red to purple colors), at the low temperature $T=t/100$.  

For both phonon frequencies considered, the imaginary part of the self energy displays the approximate step behavior expected from Eq.~\ref{eq:imsigmau0}, with a degree of broadening partly due to temperature effect and partly an artifact of analytic continuation. The real parts shown in ($b,f$) reveal the expected linear increase followed by a maximum (the logarithmic divergence is rounded by a combination of temperature and analytic continuation errors) and then a sign change, as in Eq.~\ref{eq:resigmau0}. 

Commenting specifically on the smaller phonon frequency, $\omega_0/t = 0.02$, data shown in Fig.~\ref{fig:self_energies}($a,b$), we see that because the temperature is within a factor of two of the phonon frequency, Im$\Sigma(\omega=0)$ is visibly non-zero. The increase in Im$\Sigma$ occurring as $\omega$ is increased has a magnitude of order $\lambda\omega_0$ but the onset (defined as the halfway point of the step) decreases as $\lambda$ increases. The real part shows a peak of height increasing with $\lambda$, but the frequency at which Re$\Sigma$ is maximal does not shift noticeably with $\lambda$. 

For the larger phonon frequency, $\omega_0/t = 0.2$, we see that as $\lambda$ is increased the effective phonon frequency, $\tilde{\omega}_0$, (identified from midpoint of the step height in Im$\Sigma$ or the peak of -Re$\Sigma$) decreases. Additionally, we note the step height of Im$\Sigma$ becomes larger than $\frac{\pi \lambda\tilde{\omega_0}}{2}$. For Re$\Sigma$ in ($f$), as $\lambda$ increases, the magnitude of the low frequency slope becomes larger than $\lambda$, while the peak position and point where $-\mathrm{Re}\Sigma$ crosses zero decrease corresponding to the value of $\tilde{\omega_0}$.
We also observe a weak maximum just above the step onset of Im$\Sigma$, and see fine structure appear in the frequency dependence of Im$\Sigma$, indicating the possibility of nonlinear effects in the self energy. The rapid onset of nonperturbative effects in the Holstein model is not necessarily physically relevant: many materials have an electron-phonon self energy characterized by a $\lambda\sim 1$ arising from a sum over many phonon modes each of which couples relatively weakly. 

\subsection{Self energies for $U/t=2$}
\label{sec:self-energies-U2}
In Fig.~\ref{fig:self_energies}($c,d,g,h$) we present the real and imaginary parts of the self energy computed at several electron-phonon coupling strengths for a moderately correlated ($U=2t$) Fermi liquid  at phonon frequencies $\omega_0 = 0.02t$ and $0.2t$. At $\lambda=0$ the expected Fermi liquid behavior is evident: Im$\Sigma\sim \omega^2$ with a non-zero but very small intercept arising from the $T^2$ $\omega=0$ scattering rate, and Re$\Sigma\propto \omega$ with the slope corresponding to a mass enhancement $m^\star/m\approx 1.4$. The $\lambda\neq 0$ results show that adding an electron-phonon coupling produces additional low frequency structure in the self energy.

For $\omega_0=0.2t$, the additional structure in Im$\Sigma$ (Fig.~\ref{fig:self_energies}($g$)) is an approximately step-like onset of additional scattering. Similarly in Re$\Sigma$ (Fig.~\ref{fig:self_energies}($h$)) we observe with increasing $\lambda$ an additional contribution to the low frequency slope (electron-phonon mass enhancement) followed by a peak and then a decrease below the $\lambda=0$ baseline. Comparison to Fig. ~\ref{fig:self_energies}($e,f$) shows however that much larger values of $\lambda$ are required to produce a step of a given height at $U=2t$ than at $U=0$. 

For $\omega_0=0.02t$, (Fig. ~\ref{fig:self_energies}($c,d$)) the principal effect of adding phonon scattering is an increase in the dc limit of the scattering rate. Only at $\lambda$ = 0.4974, a value approaching the value at which the metallic state begins to destabilize towards a bipolaron-localized phase, are signs of the expected phonon-frequency-related structures evident in Re and Im $\Sigma$. 

The reason for the lack of frequency-dependent structure observed at $\omega_0=0.02t$ for $U=2t$ and moderate $\lambda$ is not clear. One contribution is a temperature-dependent broadening, arising from thermal scattering. Supplementary Materials (SM) Figure 1~\cite{SI} presents the  self energies for $\omega_0=0.02$ and $\beta\cdot t = 20,50,100$ corresponding to $T/\omega_0=2.5,1,0.5$. We see that for the non-interacting ($U=0$) case the structure in $\Sigma$ is rapidly washed out with increasing $T$, with structure remaining more visible for the larger $\lambda$. SM Figure 2~\cite{SI} shows a similar but much less severe effect for $U=0$ and $\omega_0=0.2t$, confirming that the physics has to do with the ratio of temperature to phonon frequency. The temperature dependence of $\Sigma(\omega)$ for $\omega_0=0.2t$ and $U=2$ shown in SM Fig. 2 clearly reveals an extra broadening that depends on temperature and on $\lambda$, suggesting that the extra electron-electron mediated scattering acts to broaden the phonon-induced frequency dependence in a way that is more severe at smaller $\lambda$. We note, however, that the lower phonon frequency case presents difficulties for the analytic continuation because as temperature increases, there are fewer Matsubara frequency points on the frequency scales of the phonon energy. This, in combination with the very small additional effect of phonons (especially for small $\omega_0$ and small $\lambda$) argues for caution in interpreting the frequency dependence found for $\omega_0=0.02t$.

\section{Electron correlation effects on the phonon contribution to the self energy}
\label{sec:ph_renorm}

In this section we consider the extent to which it is reasonable to view the full electronic self energy of the electron-phonon coupled system as the sum of an electron-phonon coupling independent electronic background and an extra term due to phonons. In the following section we consider the possibility that the electron correlation effects are modified by the presence of phonons. Viewing phonons as an add-on is the point of view underlying the standard DFT treatment of electron-phonon coupling.  From this point of view, the main effect of interactions is to change (``renormalize") the coupling between phonons and interacting electrons.  

\begin{figure}[h!]
        \centering
        \includegraphics[width=\linewidth]{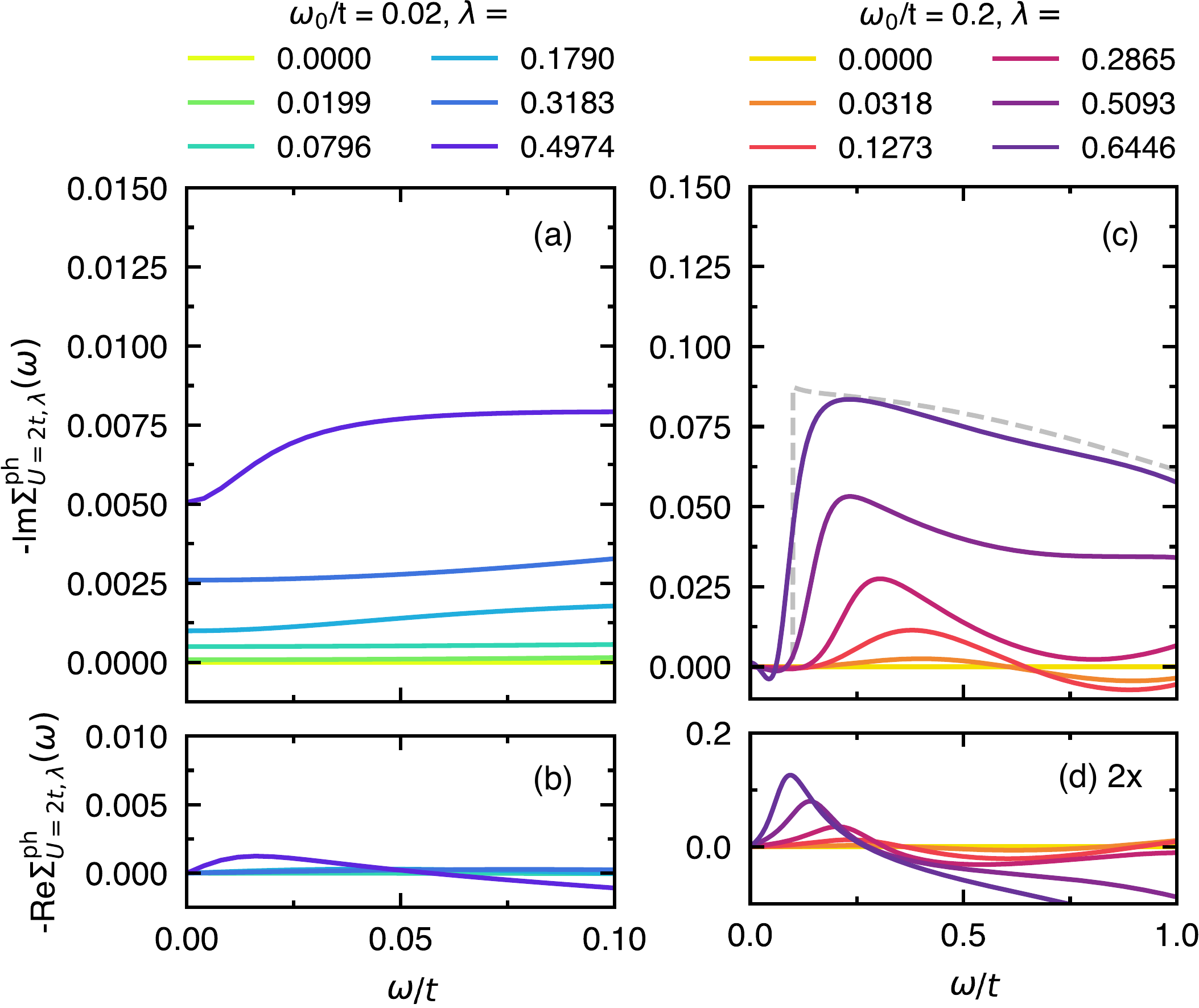}
        \caption{\textbf{Phonon contribution to the self energy for different values of $\lambda$.} Self energy differences showing the imaginary ($a,b$) and real ($c,d$) parts of the analytically continued self energies for $U/t=2$ subtracting the corresponding $U/t=2$, $\lambda = 0$ data to show the effect of phonons on the electron self energy. All data is calculated for $\beta\cdot t=100$ and values of $\lambda$ correspond to  Fig.~\ref{fig:self_energies}. In ($c$), we include a comparison to $\mathrm{Im}\Sigma(\omega) = \pi \tilde{g}^2\mathrm{A}(\omega) \theta(\omega^2 - \tilde{\omega}^2_0)$ for $\tilde{g} = g/1.5$ and $\tilde{\omega}_0 = 0.1$, plotted as a grey dashed line. Note that ($d$) shows the real part of the subtracted self energy magnified by 2x.}  
        \label{fig:self_energies_diff}
\end{figure}
To make this argument precise we plot in Fig.~\ref{fig:self_energies_diff} the phonon contribution to the electron self energy $\Sigma^{ph}_{U,\lambda}\equiv \Sigma_{U,\lambda}-\Sigma_{U,\lambda=0}$ defined as the difference between the self energy computed with $\lambda\neq0$ and with $\lambda=0$. As this quantity is the difference between two independently continued quantities, analytic continuation errors may be magnified in the difference, especially for small $\lambda$ where $\Sigma_\lambda$ is very close to $\Sigma_{\lambda=0}$ We therefore believe the larger $\lambda$ results to be more reliable. 

Comparison of the largest $\lambda$ ($\omega_0=0.02t$, $\lambda=0.4974$ and $\omega_0=0.2t$, $\lambda=0.6446$) results for Im$\Sigma^{\mathrm{ph}}$ in Fig.~\ref{fig:self_energies_diff} to the $U=0$ curves presented in Fig.~\ref{fig:self_energies} shows that indeed the phonon contribution to the $U=2t$ self energy has a very similar structure to the phonon self energy at $U=0$. However, the scale of Im$\Sigma^{\mathrm{ph}}$ is much less than the size of Im$\Sigma$ for the comparable $\lambda$ at $U=0$ in Fig.~\ref{fig:self_energies}($a,e$), indicating that the electron-electron interaction has substantially weakened the effective electron-phonon coupling. The dashed curve in Fig.~\ref{fig:self_energies_diff}($c$) shows the perturbative expression evaluated using Eq.~\ref{eq:imsigmau0} but with $\lambda=0.28$ and the bare density of states replaced by the electron spectral function computed from DMFT at $U=2t$ and $\lambda=0$. We see that the imaginary part of the self energy is in essence the perturbatively expected contribution, with a coupling constant reduced by a factor of about $2.25$ relative to the bare $\lambda$.

While the $\Sigma_{ph}$ computed at larger $\lambda$ is consistent with the notion that phonons contribute a kind of quasiparticle scattering with a renormalized coupling constant, the smaller $\lambda$ $\Sigma^{ph}$ are not completely consistent with this picture. As discussed above, for the smaller $\omega_0=0.02t$ the $\Sigma_{\mathrm{ph}}$ data for smaller-$\lambda$ values exhibits a smooth frequency dependence, with no trace of the onset of scattering at the phonon frequency. For the larger $\omega_0=0.2t$, the imaginary parts of the self energy for smaller $\lambda$ value reveal a sharper peak at the scattering onset and a marked decrease (and even a sign change) at frequencies much greater than $\omega_0$. The extent to which these behaviors arise from analytical continuation uncertainties remains to be determined.

Comparison of panel ($d$) of Fig.~\ref{fig:self_energies_diff} to panel ($f$) of Fig.~\ref{fig:self_energies} reveals that increasing the interactions shifts the zero crossing in Re$\Sigma$ to a lower frequency and increases the magnitude of the higher frequency part of Re$\Sigma$. We attribute this to phonon-induced changes in the high frequency part of the self energy, discussed in Section~\ref{fig:high_omega_SE}.

We now use characteristic features of $\Sigma^{ph}$ to define renormalized couplings $\tilde{\lambda}$ and phonon frequencies $\tilde{\omega}_0$ quantitatively. For the larger phonon frequency, we defined $\tilde{\omega}_0$ from the maximum in Re$\Sigma$ (equivalently from the midpoint of the rise in Im$\Sigma$) and then we fit the peak in Im$\Sigma$ to the perturbative form for $\Sigma$ as in Equations~\ref{eq:imsigmau0} and~\ref{eq:resigmau0} using the base density of states $N(\mu)$. For the smaller phonon frequency, the lack of frequency-dependent structure in the self energy (for most $\lambda$) makes it difficult to define a $\tilde{\omega}_0$; we instead use the intercept of the imaginary part of the self energy, which we expect to be $\tilde{\lambda}\pi T$, to define $\tilde{\lambda}$.

\begin{figure}
\centering
    \includegraphics[width=0.73\linewidth]{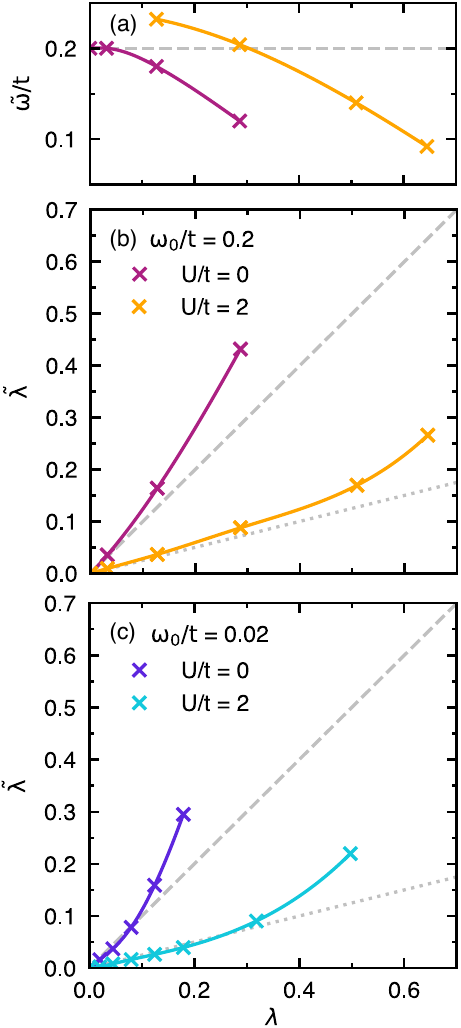}
    \caption{\textbf{Renormalization of $\lambda$ and $\omega_0$.} The renormalized phonon frequency $\omega_0$ and electron-phonon coupling strength $\tilde{\lambda}$ found for  $\omega_0/t = 0.2$ (panels $a-b$) and  $\omega_0/t = 0.02$ (panel $c$). Grey lines are to guide the eye; the dashed line shows a linear relationship between $\lambda$ and $\tilde{\lambda}$, and the dotted grey line shows $\tilde{\lambda}$ = $\frac{\lambda}{4}$.} 
    \label{fig:renorm_lambda}
\end{figure}

Panel ($a$) of Figure~\ref{fig:renorm_lambda} shows the renormalized phonon frequencies obtained as described above for $\omega_0=0.2t$ as functions of the bare coupling $\lambda$, (noting that for the $U/t=2$ case the analytic continuation issues mentioned above lead to some uncertainties for small $\lambda$ values). Also for $\omega_0/t=0.2$, ($b$) shows the extracted $\tilde{\lambda}$. For $U/t=0$, we observe that $\tilde{\lambda}$ depends nonlinearly on $\lambda$; deviations from $\tilde{\lambda}=\lambda$ (grey dashed line) become noticeable by $\lambda\approx 0.25$ at which point the coupling to electrons has softened the phonon frequency by almost a factor of two. In the correlated ($U/t=2$) case, we see that at weak coupling the effective electron-phonon coupling is reduced by approximately a factor of 4 for smaller $\lambda$ values. Deviations from the linear relation become noticeable for $\lambda\approx 0.5$ ($\tilde{\lambda}\approx 0.15-0.2$). The relative magnitude of the nonlinearity (deviation from weak coupling linear behavior) is larger for $U=2t$ than for $U=0$.

For the smaller phonon frequency, shown in Figure~\ref{fig:renorm_lambda}($c$), we see that for $U/t=0$ we see that the nonlinear correction becomes rapidly large as $\lambda$ exceeds $0.15$. For $U/t = 2$, we see again an approximately factor of 4 reduction of $\tilde{\lambda}$, again with the nonlinear effects growing more rapidly than in the $\omega_0=0.2t$ case.

We note that an alternative way to estimate the renormalization of the phonon frequency is via the phonon self energy calculated to leading order in $g^2$ as $\Pi(i\nu_n)= - g^2 \chi(i\nu_n)$; 
the $\nu=0$ value is interpreted as a shift in the phonon frequency $\tilde{\omega}_{\rm ph} = \omega_0-\Pi(\nu=0)$. We show in the Supplementary Materials~\cite{SI} that this method produces similar results to the above investigation of the electron self energy. For example, at (U/t = 2, $\omega_0/t$ = 0.2, $\lambda$ = 0.509) we find that
the peak-height extraction gives us $\tilde{\omega}_0/t$ = 0.139, while using $\chi$ gives
$\tilde{\omega}_0/t$ = 0.129. We note however that the phonon self energy method provides only a leading order (in $g^2$) shift to the phonon energy, while the above discussion using the electron self energy includes higher order effects as well. We provide a calculation and discussion of the phonon self energy calculation in the Supplementary Materials as a point of comparison~\cite{SI}.

\section{Renormalization of correlation strength due to phonons \label{sec:correlation_renorm}}

In the previous section, we analyzed the numerical results from the point of view of a phonon addition to a given electronic background. In this section we consider the converse question of how the electronic background might change due to the electron-phonon coupling. In a moderately correlated Fermi liquid at low $T$ and for low frequencies the imaginary part of the self energy is $Im\Sigma(\omega)=C\left((\pi T)^2+ \omega^2\right)$, increasing to a maximum at a frequency of order of the bandwidth and decreasing as $\omega$ is further increased. 

\subsection{Frequency dependence of the self energy}

\begin{figure}[H]
        \centering
        \includegraphics[width=0.9\linewidth]{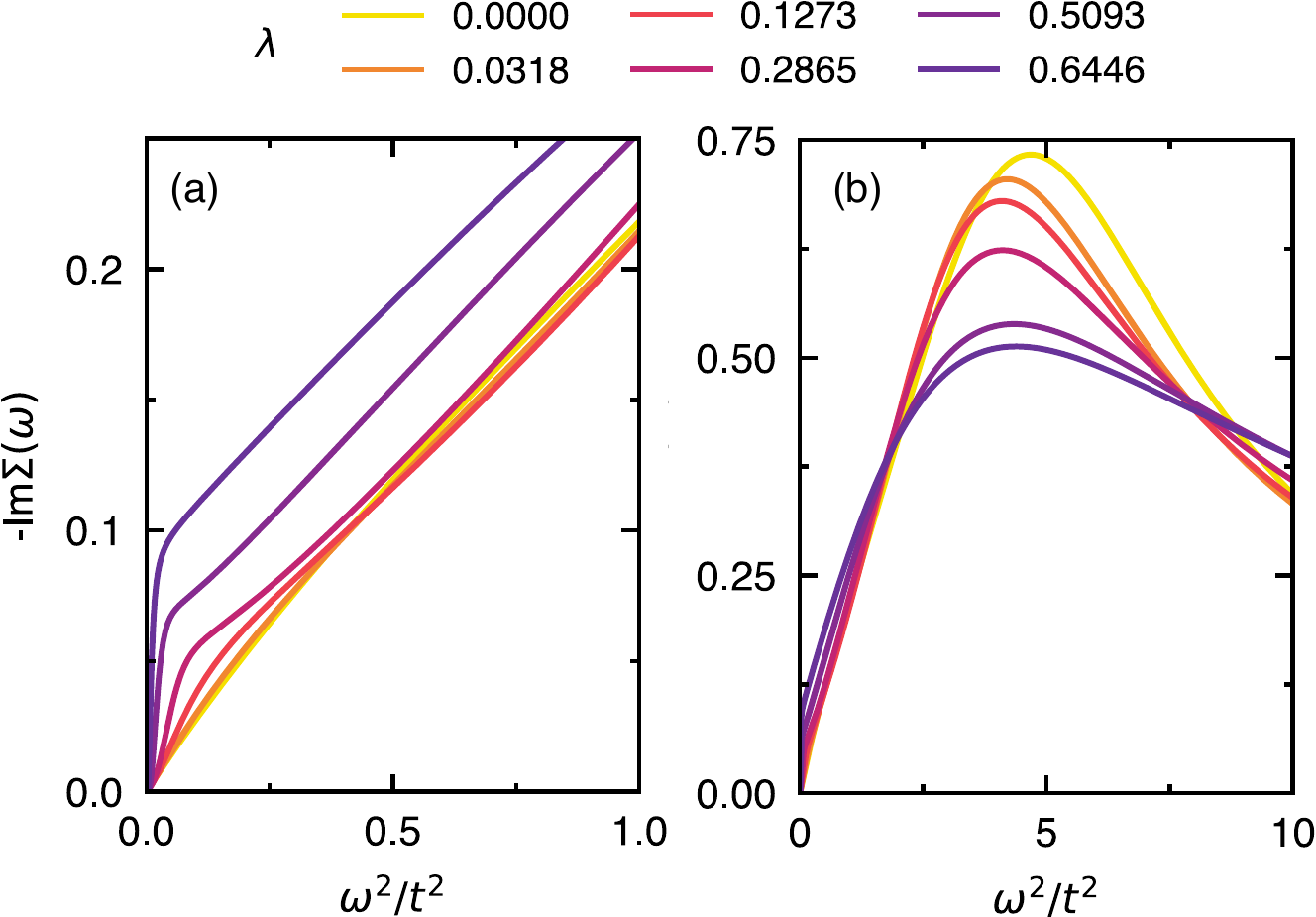}
        \caption{\textbf{High-frequency electron self energy.} The electron self energy plotted against $\omega^2$ is shown for $\omega_0/t = 0.2$ and $U/t= 2$, for ($a$) a narrower range of low-intermediate and ($b$) over a wider range extending to high frequencies. As a reference for the slope, the dashed line in ($a$) is $\frac{\omega^2}{5}$. 
         }
        \label{fig:high_omega_SE}
\end{figure}

\begin{figure*}
        \centering
        \includegraphics[width=0.9\textwidth]{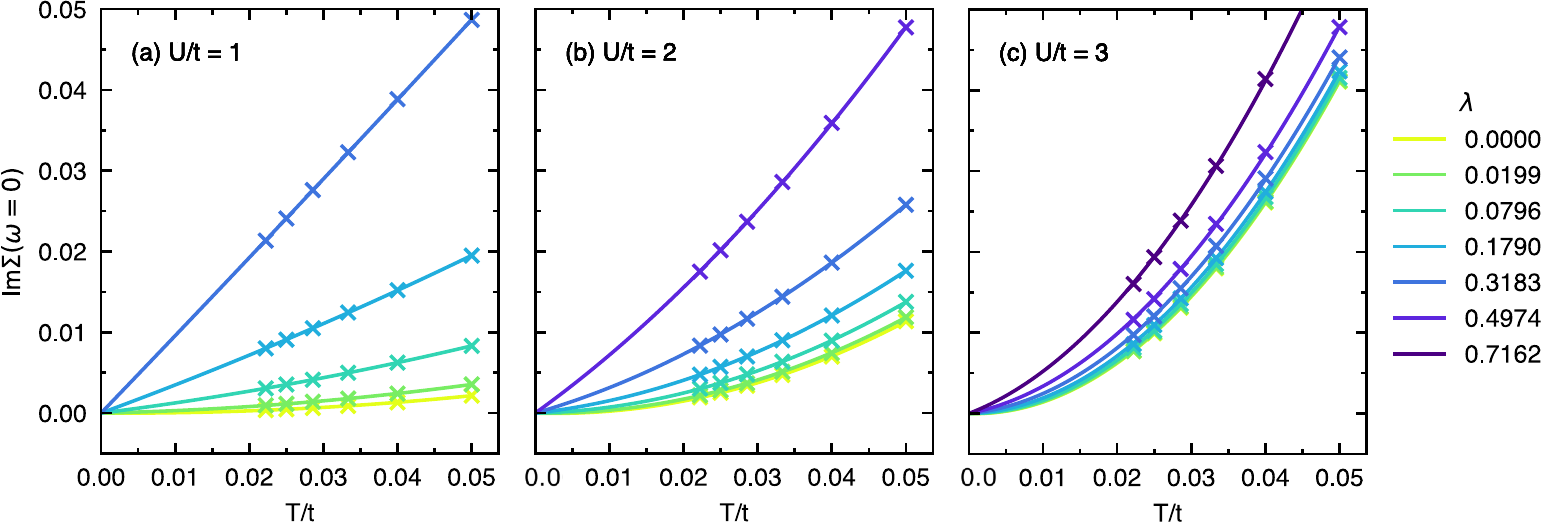}
        \caption{ 
         \textbf{Electronic decay rates due to electron-electron and electron-phonon interactions.} Im$\Sigma(\omega=0)$, the electronic decay rate, calculated by cubic spline interpolation of the Matsubara frequency axis data for $U/t$ = 1 ($a$), 2 ($b$), and 3 ($c$) for a bare phonon frequency of $\omega_0/t$ = 0.02 and increasing electron-phonon coupling strength $\lambda$ across a range of temperatures from $\beta t =$ 45 to 25. For this range, the phonon frequency is smaller than the temperature scale, resulting in a linear T dependence of Im$\Sigma(\omega)$.  
         } 
        \label{fig:scattering_rates}
\end{figure*}

Fig.~\ref{fig:high_omega_SE} plots the imaginary parts of the calculated self energies against $\omega^2$ over low-moderate frequency regime (left panel) and a broad frequency regime extending to energies greater than the half bandwidth $2t$ (right panel). Panel (a) shows that the $C$ coefficient is essentially $\lambda$-independent at approximately $C\approx1/5$, demonstrating that the electron-phonon interaction does not significantly impact the low-frequency electron-electron interactions. The low frequency limit of the real part of the self energy (SM Table 1)~\cite{SI} also shows negligible dependence on $\lambda$. 
Separating the low-frequency slope into phonon and electron-electron parts via $\frac{\mathrm{dRe}\Sigma}{\mathrm{d}\omega}=-\tilde{\lambda}-\left[\frac{\mathrm{dRe}\Sigma}{\mathrm{d}\omega}\right]_{\rm ee}$ with $\tilde{\lambda}$ taken from Fig.~\ref{fig:renorm_lambda}, we find the remaining value of -$\left[\mathrm{dRe}\Sigma/\mathrm{d}\omega \right]_{\rm ee}$ is $\lambda$-independent with the exception of very high $\lambda$. In other words, the electron mass enhancement and scattering rate that define the low-frequency behavior of the correlated Fermi liquid state are only very weakly affected by the electron-phonon coupling. 

However, from panel ($b$) we see that the higher frequency scattering rate ($\omega\gtrsim 2t$) is noticeably affected. We note that the analytical continuation methods used are less reliable in this range, but the magnitude and systematic dependence of the effects on the electron-phonon coupling strength suggests that the dependence may be qualitatively correct. Additionally, calculations of $A(\omega) = \frac{-\mathrm{Im}G(\omega)}{\pi}$ presented in the SM~\cite{SI} show that changes in $A(\omega)$ alone are too small to explain this high-frequency trend. This suggests the extra scattering provided by the electron-phonon effects broadens the band edge and yields an extra decay channel at high frequencies. This is in agreement with the findings of Ref.~\cite{sangiovanni_electron-phonon_2005}; that phonon effects which cannot be described by a rescaling of electron-phonon or electron-electron effects can appear in the high frequency features of the self energy. 
\subsection{\label{sec:scattering_rates}Decay rates}

Similarly, we can inspect the $C$ coefficient by investigating the scaling of the scattering rate as a function of temperature. Fig.~\ref{fig:scattering_rates} shows the T-dependent decay rates (Im$\Sigma(\omega\rightarrow 0)$) from the $\omega_0/t = 0.02$ data, extracted using a cubic spline interpolation to extrapolate our Im$\Sigma(i\omega_n)$ data to zero Matsubara frequency. This figure contains data in a temperature range of $\beta t$ = 25 to 45, as for these values $T>\omega_0$, resulting in a $T$-linear trend as expected for scattering by classical phonons. Through a Im$\Sigma(\omega=0) = \alpha T + C T^2$ fit of the data, contributions from electron-phonon and electron-electron interactions can be analyzed, with $C$ coefficients presented here in Table 1 and both $C$ and $\alpha$ shown in the SI.
\begin{table}[H]
\begin{center}
\begin{tabular}{| c |c|c|c|}
\hline 
\ \ \ \ \ $\lambda$ \ \ \ \ \ & \ U/t = 1 \   & \ U/t = 2 \ & \ U/t = 3 \ \\ \hline  \hline 
0.0000 	&	0.9814 	&	5.097	&	17.020	\\	
0.0199	&	1.0264	&	5.021	&	16.936	\\	
0.0796	&	1.0186	&	5.038	&	16.763	\\	
0.1791	&	1.0420	&	4.968	&	16.409	\\	
0.3183	&	0.3552	&	5.015	&	15.918	\\	
0.4974	&		    &	5.888	&	15.528	\\	
0.7162	&		    &		    &	16.915	\\	\hline
\end{tabular}
\label{table:slopes}
\caption{Quadratic ($C$) fit coefficients for the decay rates plotted in Figure~\ref{fig:scattering_rates} for increasing values of $U$ and $\lambda$ with phonon frequency $\omega_0/t = 0.02$.}
\end{center}
\end{table}
The panels of Fig.~\ref{fig:scattering_rates} present scattering rates for $\omega_0/t = 0.02$ calculated using ($a$) $U/t = 1$, ($b$) $U/t = 2$ and ($c$) $U/t = 3$. As expected from earlier sections, a comparison of panels ($a-c$) and the $\alpha$ values in the SM~\cite{SI} clearly shows that the linear electron-phonon contribution to the decay rate is reduced in the presence of even weak correlations. Of significance to this section, from the fit coefficients reported in Table 1 we can see that for all values of $U$ in Fig.~\ref{fig:scattering_rates}, and that the $T^2$ coefficient of the fit is almost unchanged by the value of $\lambda$ until again we reach the largest values studied, those approaching an energy scale to compete with $U$. This reinforces the finding of Section~\ref{sec:correlation_renorm}, that for moderate electron-phonon coupling strengths the correlation effects will be largely unaffected by coupling to phonons. 

\section{\label{sec:conclusion}Conclusion}

The current mainstream theoretical approach to calculating electron-phonon coupling effects in solids is based on a density functional description of electronic and lattice energetics and couplings combined with a one-loop treatment of the effects of the electron-phonon coupling. Generalizing this formalism to the case of correlated materials with electronic properties not adequately described by density functional theory is an important goal of materials theory. A basic result of the mainstream theory is that phonon effects are in essence a perturbative ``add-on" to the no-phonon physics, leading to carrier scattering  and low frequency mass renormalization but not changing the fundamental physics of the state; a natural question is whether this is the case in correlated quantum materials.

As a step towards the development of a beyond DFT electron-phonon theory, this paper investigates the effects of beyond-DFT and beyond one-loop physics via a comprehensive non-perturbative study of the Hubbard-Holstein model. Our computations of the frequency and temperature dependence of the electron self energy over wide frequency and temperature ranges and for a variety of model parameters enables us to draw two important conclusions: first, that low-frequency signatures of electronic correlations such as the electronic component of the mass enhancement and $T^2$ and $\omega^2$ contributions to the scattering rate and are only weakly impacted by coupling to phonons, with the exception of very large electron-phonon coupling strengths (at frequencies of order the bandwidth the effects are larger); and second, that as in the conventional theory, electron-phonon coupling leads to a perturbative add-on effect to the self energy, whose amplitude may be quantified by a renormalized electron-phonon coupling. Therefore, we use the magnitude of the phonon contribution to the self energy to define a renormalized electron-phonon coupling which we find to be dramatically suppressed by even moderate electronic correlations.

This finding is consistent with an important earlier paper of Huang, Hanke, Arrigoni and Scalapino who used quantum Monte Carlo methods to compute the vertex correction diagram at zero Matsubara frequency for the bare electron-phonon coupling of the Holstein-Hubbard model~\cite{Huang03}, finding that the vertex indeed suppressed the coupling. Direct comparison is difficult because their coupling is defined in a Fermi liquid sense as a ratio of a vertex to a self energy and the methods available at the time limited their work to high temperatures $T=t/2$,  $8\times8$ lattices, and to leading order in the electron-phonon coupling. Possible phonon-induced renormalization of the electron-electron interaction were not considered.

Our work builds on a large previous literature~\cite{Huang03,koch2004renormalization,rosch2004apparent,sangiovanni_electron-phonon_2005, sangiovanni2006electron,gunnarsson2008interplay,moghadas2025effective} on electron-phonon interactions in a variety of model systems. These works demonstrated an interesting interplay between electron-electron and electron-phonon coupling in particular showing (in the case of the Holstein Hubbard model) that phonon-induced changes to the electron mass enhancement interactions are suppressed at low to moderate correlation strengths but can be enhanced near a Mott transition~\cite{moghadas2025effective, sangiovanni_electron-phonon_2005, koch2004renormalization}. Where there is overlap, our results agree with previous works~\cite{sangiovanni_electron-phonon_2005, sangiovanni2006electron, moghadas2025effective, Huang03, koch2004renormalization}, for example, in qualitative agreement with respect to the suppression of electron-phonon coupling strength for Holstein like phonons as in Refs.~\cite{Huang03,sangiovanni2006electron}, and in terms of higher frequency self energy changes due to electron-phonon interactions as in Ref.~\cite{sangiovanni_electron-phonon_2005}.  However, in the prior literature the variation of the self energy as a function of frequency was not much studied and it was not settled whether the weak dependence of the electronic mass enhancement on the strength of the phonon coupling were due to an interplay between a phonon-induced suppression of the electronic contribution and an extra phonon contribution.

We also note the work of R\"osch and Gunnarson~\cite{rosch2004apparent, gunnarsson2008interplay}  who applied sum rule arguments to a t-j model to investigate the effect of correlations on electron-phonon coupling and found that electron-electron interactions reduce the a measure of the overall magnitude of the phonon but not the electron self energies. While a direct comparison is not possible because we study the Hubbard rather than t-J model, the strong difference we find between the low and high frequency effects on the electron self energy means that results of sum rules, which integrate over all frequencies, should be interpreted with caution.

Our work lends support to the notion that, even in the correlated case, the low (relative to bandwidth) frequency effects of electron-phonon coupling can be treated perturbatively, but with a coupling strongly renormalized with respect to the DFT value. This implies that  the materials theory question is to find a workable method of determing the coupling constant renormalization. Our results may be useful as a benchmark for density functional plus dynamical mean field calculations of coupling constant renormalizations in real materials contexts.  Extension of our results models involving other electron-phonon coupling mechanisms (e.g. the Jahn-Teller/orbital splitting phonons relevant in materials with degenerate partly filled d-shells and the gradient couplings relevant to long wavelength phonons) is an important topic for future investigation. 

Further examination of the limits to our results is also important.  Studies of alternative models suggest that larger phonon frequencies~\cite{lau2025oscillate} or spin-phonon couplings~\cite{hardy2025enhanced} may produce interesting changes to correlation effects, meaning that in these situations computations of a renormalized coupling might not suffice. Extending our analysis to lightly doped Mott insulators, and models in the vicinity of a Mott transition (where among other things electronic energy scales may be small) or to  multi-orbital models are important goals. Finally, our results indicate that correlation effects may significantly modify the predictions of the current DFT-based framework, motivating further investigation of electron-phonon coupling in beyond-DFT electronic structure methods. 
\\ \\
\section*{DATA AVAILABILITY} 
The data that support the findings of this article are avail-
able from the author upon reasonable request. The CTSEG code from the TRIQS software suite is available at \url{https://github.com/TRIQS/ctseg}.  
\\ \\
\section*{Acknowledgments} The authors thank D. Abramovitch, A. Hardy, A. Hampel, O. Parcollet, and A. Poli for many helpful discussions. The Flatiron Institute is a division of the Simons Foundation. 

\bibliography{bib.bib}

\end{document}